\pdfoutput=1
\documentclass[aps,prl,twocolumn,showpacs,superscriptaddress,groupedaddress]{revtex4-1}  

\usepackage{graphicx}  
\usepackage{dcolumn}   
\usepackage{bm}        
\usepackage{amssymb}   
\usepackage{color}
\usepackage{verbatim}
\usepackage{enumerate}
\begin{document}

\widetext

\title{Structural Responses of Quasi-2D Colloid Fluids to Excitations Elicited by Nonequilibrium Perturbations}
                            
\author{Jelena Pesic}
\thanks{These authors contributed equally.}
\author{Joseph Zsolt Terdik}
\thanks{These authors contributed equally.}
\author{Xinliang Xu}
\author{Ye Tian}
\author{Alejandro Lopez}
\author{Stuart A. Rice}
\author{Aaron R. Dinner}
\author{Norbert F. Scherer}
\affiliation{Department of Chemistry and James Franck Institute, The University of Chicago, 929 E.\ 57th St., Chicago, IL 60637}

\date{\today}

\begin{abstract}
We investigate the response of  a dense monodisperse quasi-two-dimensional (q2D) colloid suspension when a particle is dragged by a constant velocity optical trap. Consistent with microrheological studies of other geometries, the perturbation induces  a leading density wave and trailing wake. We also use a hybrid version of Stokesian Dynamics (SD) simulations  to parse direct colloid-colloid and hydrodynamic interactions.  We go on to analyze the underlying individual particle-particle collisions in the experimental images.  The displacements of particles form chains reminiscent of stress propagation in sheared granular materials.  From these data, we can reconstruct steady-state dipolar-like flow patterns that were predicted for dilute suspensions and previously observed in granular analogs to our system.  The decay of this field differs, however, from point Stokeslet calculations, indicating that the non-zero size of the colloids is important.  Moreover, there is a pronounced angular dependence that corresponds to the surrounding colloid structure, which develops in response to the perturbation.  Put together, our results show that the response of the complex fluid is highly anisotropic owing to the fact that the effects of the perturbation propagate through the structured medium via chains of colloid-colloid collisions.
\end{abstract}

\pacs{83.50.-v, 82.70.-y, 83.10.-y}
\maketitle
\section{ Introduction}
Provided that one neglects inelastic collisions between particles, all fluids should develop hydrodynamic-like dynamics on a sufficiently large time scale. 
Hence, it is reasonable to expect that the same, or at least analogous, principles will govern all forms of complex fluids. To the extent that this is the case, it is important to determine how the average material response emerges from the underlying mechanisms of force transmission and dissipation.
The extreme cases are simple fluids, for which the hydrodynamic description is well established \cite{khair2006single}, and athermal granular bead packs, for which structural characterizations exist \cite{liu1995force, jaeger1996granular, o2001force, majmudar2005contact} but descriptions of the dynamics remain limited \cite{kadanoff1999built, grossman1997towards, dufty2008linear, weeber2011hydrodynamic}.

Recent theoretical \cite{gazuz2009active,squires2008nonlinear} and experimental \cite{habdas2004forced,sriram2010active} advances now allow the study of microscopic force propagation in complex fluids, such as glasses, colloid suspensions \cite{gazuz2009active}, polymers \cite{witten2010structured}, and granular materials \cite{xu2009equivalence}.  In such experiments \cite{habdas2004forced, waigh2005microrheology,habdas2002video}, optical or magnetic tweezers coupled with optical microscopy are used to micromanipulate  a probe particle to deliver a precise and localized mechanical perturbation to a system while simultaneously detecting the system response.  In this fashion, microviscosities in shear thinning and thickening regimes have been determined as functions of the speed of the probe particle and related to the average structure of the fluid around it \cite{sriram2009small, sriram2010active, wilson2009passive}.

Dense colloid fluids, in which thermal motion and hydrodynamic interactions both play significant roles, lie between the liquid and granular regimes. Very recent work on sheared colloid fluids \cite{bradyPhysToday, cohen2011shear,schall2010shear}  draws parallels to experiments on sheared granular packs \cite{liu1995force} while also highlighting the complex interplay between granular and liquid components. Sheared dense colloid fluids exhibit a non-monotonic response with applied shear force, as exemplified by shear thinning and shear thickening.  The recent advances of microrheological methods now make it possible to investigate the relationship between these non-Newtonian responses and changes in the structure of the sheared fluid.  We expect changes in structure to play an important role in strongly confined fluids because the response of a confined fluid to mechanical perturbation combines the effects of excluded volume and wall constraints.  This reasoning leads to the question of whether the viscosity inferred from the force required to push a bead inside a microscopic cell via Stokes' Law is appropriate if, for example, the motion of the bead alters the local structure of the cell.
 
As a step towards establishing the structure and dynamics during the response of a complex fluid to strong mechanical perturbation, and examining the role of individual particle-particle collisions, we used localized perturbations to probe the complex properties associated with dense colloid fluids. The system that we study is a quasi-two dimensional (q2D) monodisperse colloid suspension.  
We term it q2D for two reasons.  
First, by construction the experimental chamber has a thickness only slightly greater than one colloid particle diameter \cite{cui2001dynamical}, thereby restricting the centers of the colloid particles to lie close to a plane but allowing small amplitude deviations from that plane.  
Second, the supporting liquid in the colloid suspension moves in three dimensions, and the functional form, sign, temporal decay, and density dependence of the effective colloid-colloid interactions are affected by the boundary mediated transfer of momentum and mass in the liquid \cite{diamant2009hydrodynamic, cui2004anomalous, diamant2005hydrodynamic}.
We note that at the densities of our experiments, the system is in the hexatic phase. 

First we show that the structure of the fluid around one particle driven steadily by an optical trap reprises previous observations and calculations \cite{meyer2006laser,foss2000structure,wilson2011microrheology}.  
Specifically, we observe a density wave in front of, and a trailing wake following, the probe particle.
Surprisingly, our simulations suggest that hydrodynamic interactions are not required to obtain the density wave and trailing wake, although the fluid does influence the size of the wake by setting the self-diffusion constant of the colloid particles.  
Second, the angle-resolved pair correlation function exhibits a prominent, and heretofore not observed, banding structure trailing the probe particle. 
The inter-band spacing is consistent with a partial structural rearrangement of the colloid fluid from hexagonal toward square packing. 
Third, we characterize structurally anisotropic displacement chains (excitations of the system) that precede the probe particle. 
We show that the orientation of the hexatic order in our dense q2D systems influences the formation of these displacement chains. 
Finally, analysis of individual colloid particle displacements reveals a flow field around the driven particle with closed circulation loops up to 10 particle diameters from the probe particle.
The high particle density used in our experiment induces long range spatial correlations that are evident in the observed flow field. 
Hence, the decay of this field deviates from the $r^{-2}$ behavior predicted for dipolar flow using the Stokeslet approximation in q2D systems \cite{diamant2009hydrodynamic, cui2004anomalous, diamant2005hydrodynamic}.  
Furthermore, the average displacement field shows wave-like modulation preceding the probe particle, with peaks occurring at integral multiples of the particle diameter, consistent with the formation of displacement chains. Clearly, the fluid is profoundly affected by the structure that develops from individual particle-particle collisions.

\section{Methods}

\subsection{Experimental System}
\begin{figure}
\includegraphics[width=0.5\textwidth]{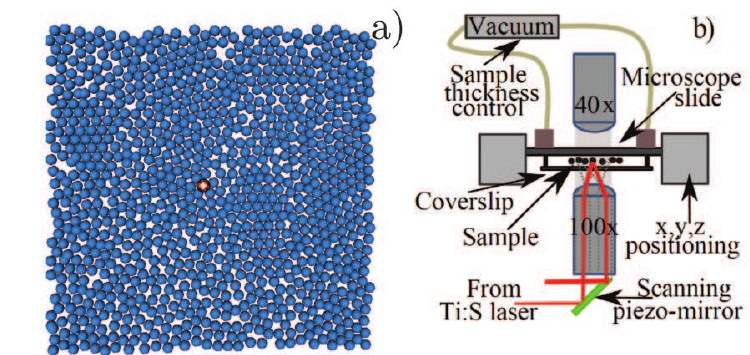}
\caption{(Color online)
 Experimental setup. (a) Typical configuration of q2D monolayer of 2.56 $\mu$m diameter silica beads with the probe particle at the center in red with a white cross. The field of view is $82 \mu\textrm{m}\times 82\mu\textrm{m}$ and contains approximately 1000 particles. (b) Optical trapping setup and thickness-controlled sample chamber. 
\label{exptSetup}
}
\end{figure}

The experimental system that we have studied is a monodisperse colloid suspension confined between hard walls (Fig. \ref{exptSetup}a)\cite{jelenaThesis}.  The colloid fluid consists of uniform size silica spheres (diameter   $\sigma$= 2.56 $\pm$ 0.04 $\mu$m; Bangs Laboratories) suspended in nanopure water and confined to a chamber with 3.2 $\mu$m wall separation maintained at $23^\circ$C.  The chamber fabrication was adopted from \cite{cui2001dynamical}. The perpendicular confinement is achieved by modulating the hydrostatic pressure inside the sealed chamber using a hand pump. The height of the sample cell is determined by measuring the distance traversed by a closed loop piezo-electric stage between the epi-reflections of a reference laser (532 nm) from the top and bottom glass surfaces of the sample cell. The chamber walls are coated with a non-wetting hydrophobic layer to prevent adsorption of colloid particles. Due to the effective repulsion between a colloid particle and the walls, the colloid particles are confined to the center plane of the chamber and form a monolayer with a height distribution of width $\pm$50 nm.  A 810 nm laser beam that is tightly focused with a high N.A. objective (Olympus SAPO 100x oil immersion) forms an optical trap (Fig.\ \ref{exptSetup}b) with a force constant $\kappa$ = 0.2 pN/nm in the sample plane. This trap strength is sufficient to overcome the force threshold (approximately $40 k_BT/\sigma$) associated with breaking through the cage of first neighbors at close packing \cite{gazuz2009active}. All of the experiments were carried out in dense colloid suspensions; the sample was translated at constant velocity past the trapped probe particle (i.e., constant dragging). The driving speed ranged from 4.9 to 12.6 $\mu \textrm{m/s}$.  The total field of view ($L\times L$) was $82\times 82\;\mu\textrm{m}^2$ and the packing fraction, $\rho\equiv N\pi \sigma^2/4L^2\approx0.76$. Examination of the bond orientational order and the translational order reveals that at this density the system is in the hexatic state \cite{jelenaThesis, marcus1997Hexatic}. 

\subsection{Analysis}

\begin{table}[htdp]
\caption{Summary of experimental parameters}
\begin{center}\begin{tabular}{|c|c|}\hline Drag Speed, $V$, ($\mu$m/s) & P\'eclet Number, $Pe$ \\\hline 12.6 & 190 \\\hline 8.3 & 125 \\\hline 6.5 & 100 \\\hline 4.9 & 74 \\\hline\hline\hline  Particle Diameter ($\sigma$)  & 2.56 $\pm$ 0.04 $\mu$m\\ \hline  Average Packing Fraction ($\rho$)  &  $\approx 0.76$ \\ \hline Field of View ($L^2$) & $82\times82\;\mu\textrm{m}^2$ \\ \hline Video Frame Rate & 67 Hz \\ \hline \end{tabular} 
\label{exptDrag}
\end{center}

\end{table}

We characterized the strength of the perturbation generated by dragging a particle by the P\'eclet number, $Pe=\sigma V/D$, where $D$ is the single particle diffusion coefficient at infinite dilution (which we estimate from the Stokes-Einstein equation for spheres in 3D), $V$ is the velocity of the dragged particle and $\sigma$ is the particle diameter. 
Physically, $Pe$ is the ratio of amplitudes of driven motion to thermal induced Brownian motion for an isolated particle. Here we reference the P\'eclet number with respect to the diffusion coefficient for an isolated particle in order to present a consistent measure of the strength of the perturbation which is independent of local density fluctuations around the probe particle. The particle positions and trajectories are extracted from the experimental data using commercial particle tracking software (DiaTrack). The centers of the particles are determined to $\sim$10 nm spatial resolution. Table \ref{exptDrag} summarizes the important experimental parameters.

\subsection{Stokesian Dynamics simulations}

A key characteristic of a colloid suspension is the discrepancy in size and mass between the colloid particles and the molecules of the supporting liquid.  
The natural separation of length and time scales in such a system implies that the momentum relaxation of the supporting liquid is complete on the time scale of the colloid particle motion. As a result, the distribution of particles can be described by the Smoluchowski equation. A second implication of the natural separation of length and time scales is that the coupling between colloid particles can be decomposed as (i) a perturbation of the hydrodynamic velocity field generated by a single particle, and (ii) the corresponding back action of the perturbed velocity field on the generating particle.

Analytical calculation of the effect of the hydrodynamic coupling on the colloid diffusive motions in a dense system is very complicated \cite{Bhattacharya2005,Bhattacharya2006}. Consequently, we use a hybrid version of Stokesian Dynamics (SD) simulations to evaluate the contributions of hydrodynamic interactions. In such a simulation, the colloid particle configuration space trajectories are composed of successive displacements, each for a short time step $\Delta t$. The influences of particle-particle and particle-wall hydrodynamic interactions are incorporated via the values of the diffusion tensor  $\mathbf{D}$ with elements $D_{ij}$.  Specifically, the displacement of particle $i$, $\Delta x_i(t)$, is determined from the Langevin equation:
\begin{equation}
\Delta x_i(\Delta t) =\sum_j \frac{\partial D_{ij}}{\partial x_j}\Delta t + \sum_j \frac{D_{ij}F_j}{kT}\Delta t + R_i(\Delta t).
\end{equation}
$R_i(t)$ is a random displacement of particle $i$ satisfying a multivariate Gaussian distribution with zero average and variance-covariance 
\begin{equation}\langle R_i(\Delta t)R_j(\Delta t)\rangle = 2D_{ij}\Delta t.\end{equation}
$F_i$ is the sum of inter-particle and external forces acting on particle $i$ and all variables are evaluated at the beginning of the time step. 

In principle,
for a particular colloid particle configuration the diffusion coefficient, $D_{ij}$, at time $t$
 can be obtained by solving the Navier-Stokes equation with boundary conditions supplied by the particle spatial configuration at the beginning of the time step.  We used, instead, a surrogate procedure that is equivalent.  This procedure, pioneered by Brady and coworkers  \cite{ermak1978brownian,bossis1987self}, captures the effect of the hydrodynamic coupling on the diffusive motions of the colloid particles yet greatly simplifies simulation.  In this approach the hydrodynamics are captured in the presumed known diagonal and off-diagonal diffusion coefficients $D_{ii}$  and $D_{ij}$ that appear in the Smoluchowski equation.  We term our approach \textit{hybrid} Stokesian Dynamics in that the diffusion coefficients are derived from experimental data. In an earlier study of diffusion in a dense q2D colloid suspension \cite{cui2004anomalous} we evaluated the center of mass and relative pair diffusion coefficients, from which the density and separation dependences of the diagonal and off-diagonal diffusion coefficients can be determined.  A fit to those data shows that the diagonal and off-diagonal diffusion coefficients can be well approximated by  $D_{ii} =\mathbf{\alpha}$ and  $D_{ij}=\beta/r^2$ for particles with center of mass separation $r$, and that both $\alpha$ and $\beta$ are constants depending only on the wall separation and the colloid concentration.  In our hybrid Stokesian Dynamics simulations, in accord with the conditions appropriate to the experiments reported in this paper, these constants are taken to be  $\alpha =1 $ and $\beta=0.3$. 

At the beginning of each time step, the diffusion coefficients $D_{ii}$ and $D_{ij}$ were evaluated for the existing particle distribution. With these diffusion coefficients, the $N$ displacements  $\{R_i\}$ were calculated using a multivariate normal generator applying the Rotational Method described in reference \cite{barr1972comparison}.  
This method utilizes the fact that the  $k\times k$ variance-covariance matrix $\mathbf{\Sigma}$ is positive definite and symmetric.  
Then, if we generate $k$ independent univariate random variables $\mathbf{Y}=(Y_1,Y_2,\cdots,Y_k)$, we obtain a random sample of $\mathbf{X}=\mathbf{Y}\mathbf{P^{-1}}$, which has a $k$-dimensional normal distribution with zero mean and variance-covariance matrix $\mathbf{\Sigma}$ satisfying $\mathbf{P}^{'}\mathbf{\Sigma}\mathbf{P} =\mathbf{I}$, where $\mathbf{P}^{'}$
is the transpose of $\mathbf{P}$.

It is important to make two further comments concerning the procedure used to calculate $D_{ii}$ and $D_{ij}$.  
First, because they are drawn from experimental data \cite{diamant2005hydrodynamic, diamant2005correlated}, the values of $D_{ii}$ and $D_{ij}$, calculated in the fashion described above, incorporate the effects of both far-field hydrodynamic and lubrication forces. 
We note, however, that at the densities used in the earlier experiments lubrication forces were found to be insignificant \cite{diamant2005hydrodynamic, diamant2005correlated}.  
Second, we hypothesize that the particular forms quoted for $D_{ii}$ and $D_{ij}$, which are valid at packing fraction up to 0.49, remain valid at the higher packing fraction of the experiments reported in this paper. 
This idea is consistent with a result obtained from the point Stokeslet representation of the hydrodynamics. 
Namely, because of the no-slip condition at the boundaries that define the q2D chamber thickness, the three particle hydrodynamic interaction vanishes, although higher order interactions do not \cite{cui2004anomalous}.  
We discuss this issue further when comparing with the experimental results.

Our Stokesian Dynamics simulation sample consists of $N$ = 85 disks of diameter $\sigma$ in a 10$\sigma\times10\sigma$ two-dimensional box with periodic boundary conditions.  
The particle-particle interactions were taken to have the continuous but nearly hard disk form:
\begin{equation}\frac{U(r^*)}{k_BT} = C(r^*-0.5)^{-\gamma},\end{equation}  
with $r^*=r/\sigma$, $C=2\times 10^{-19}$, and $\gamma=64$.
This form for the particle-particle interaction has continuous derivatives, which simplifies the treatment of forces; it has been used previously in simulations of dense colloid systems \cite{zangi2003}. 
The system is first brought to equilibrium using a Monte Carlo procedure.  
Then, to simulate the experimental arrangement in which a moving laser trap is used to exert force on a particle, we drag one of the particles in the initially equilibrated sample with a harmonic potential whose center moves at a constant velocity $V$. 
 The interaction between the dragged particle and the center of the moving potential is described by Hooke's law, $F=-\kappa r$, where $r$ is the separation of the center of the dragged particle from the center of the moving potential and $\kappa=5000 k_BT/\sigma^2$. 
This corresponds to a trap constant approximately 70 times softer than the trap strength used in the experimental system.  
This choice facilitates numerical integration by allowing a larger time step but does not significantly impact the comparison to experiment because the deviations from the trap center are small relative to the particle diameter in both cases.
For the simulations, we use units such that $\sigma = 1$ and $D_{ii}=1$; time is then measured in terms of $\sigma^2/D_{ii}$.
The particle was dragged for 200,000 time steps of $\Delta t=5\times10^{-5}$, after which 100,000 time steps were taken to allow the system to relax back to equilibrium. 

\section{Results and Discussion}

\subsection{Average structure around the probe particle}

\begin{figure}[t!]
\includegraphics[width=0.5\textwidth]{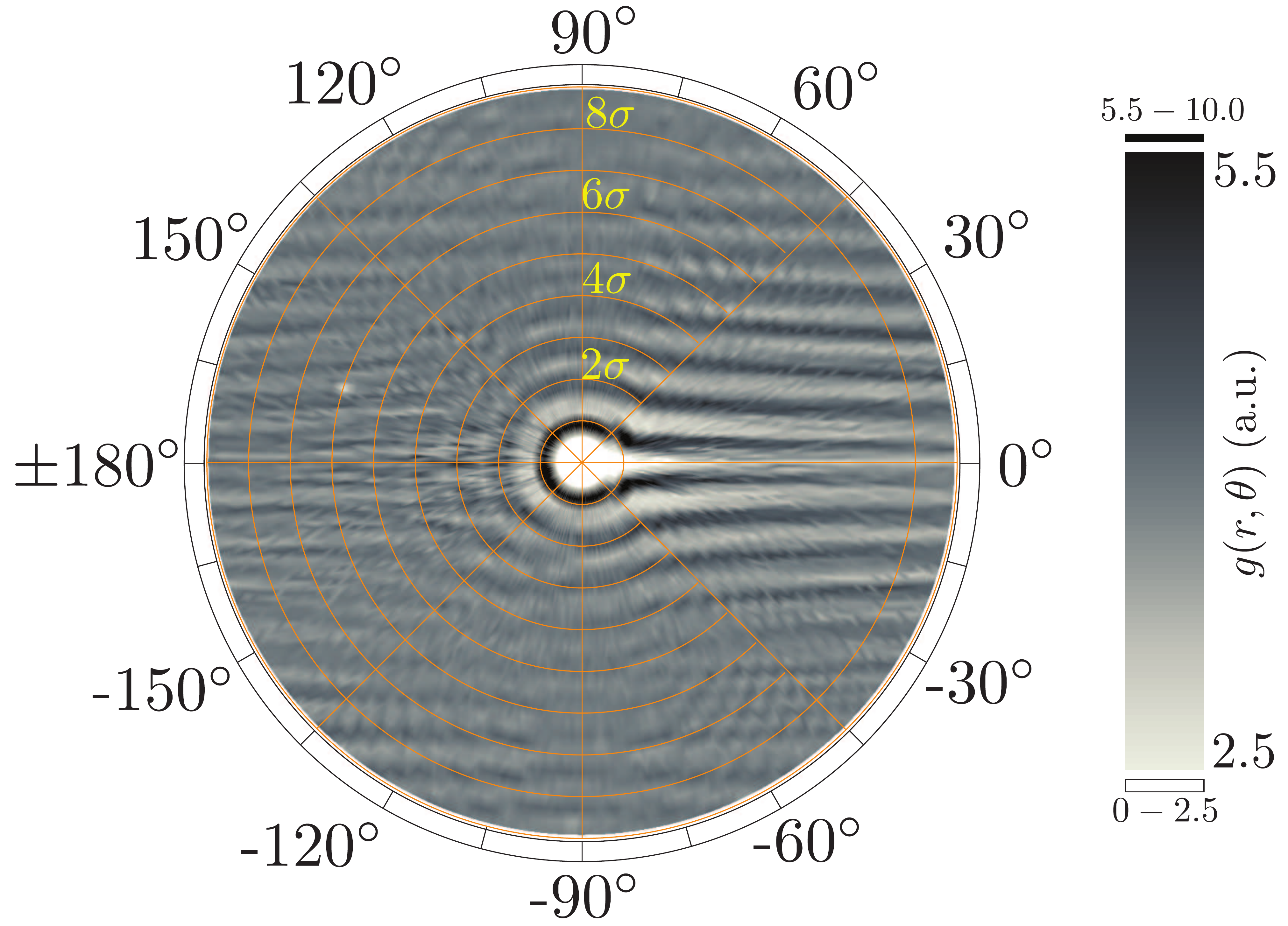}
\caption{
\label{g2Pe20Large}
(Color online)
Pair correlation function $g_2(r,\theta)$ for $Pe = 125$ centered on the probe particle, where $r$ is the inter-particle distance and $\theta$ is the angle between the flow direction (i.e., dragged particle moving to the left) and the line connecting the particle centers. The grayscale has been chosen to highlight the average structure outside the first shell. The extreme high (5.5-10) and low (0-2.5) values of the pair correlation function occur only in the first shell and the wake respectively.}
\end{figure}

\begin{figure}[t!]
\includegraphics[width=0.5\textwidth]{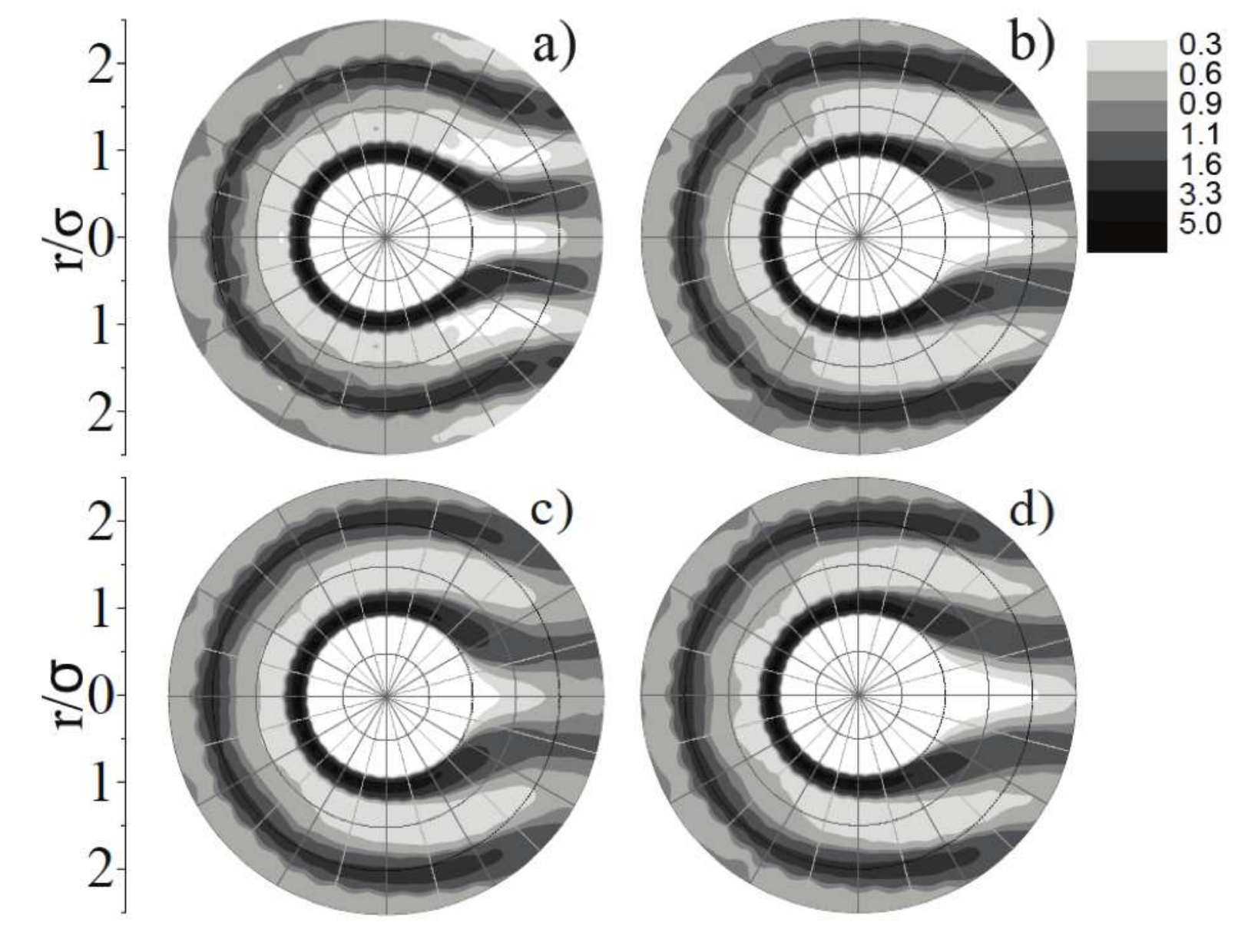}
\caption{ Pair correlation function $g_2(r,\theta)$ centered on the trapped particle, where $r$ is the inter-particle distance and $\theta$ is the angle between the flow direction and the line connecting the two particles. The experimental result for $Pe = 125$ (a) is closely reproduced by SD simulations with full hydrodynamic interactions (HI). (b). If the HI are turned off and the drag speed kept the same (c), the wake diminishes. If the HI are turned off and the velocity is increased in order to restore $\sigma V/D_{ii}$ (d), the amplitude of the wake is comparable to (b).
\label{g2SimulExptSmall}
}
\end{figure}

We first consider the change in average structure induced by a probe particle moving at constant velocity relative to the bath.  Fig.\ \ref{g2Pe20Large} displays the 2D pair correlation function, $g_2(r,\theta)$, defined with respect to the driven particle at the origin. The angle  defines the direction of the projected line between particle centers relative to the dragging direction.  We use polar binning with radial resolution $\delta r = \sigma/4$ and angular resolution $\delta \theta = 1^\circ$. The definition of the pair correlation function in both the experiment and the simulation is:
\[g(\mathbf{r}) \equiv \frac{1}{A(\mathbf{r})}\sum_{t=1}^T \sum_{i=1}^N \delta[\mathbf{r} - \mathbf{r}_i(t)],\]
where $\mathbf{r}_i(t)$ is the position of particle $i$ at time $t$ and $A(\mathbf{r})$ is the area of the bin at position $\mathbf{r}\equiv(r,\theta)$ relative to the probe particle. In other words, for all $T$ frames we bin the positions of all $N$ particles using polar coordinates and divide the number of counts in each bin by its area. The probe, located by definition at the origin, is not counted. 

The trapped particle motion generates a density wave that precedes it and a rarefaction in its wake. These features have been previously observed in experiments in unconstrained (3D) colloid systems \cite{meyer2006laser,sriram2010active} and in 2D simulations \cite{squires2008nonlinear}.  All experimental results shown are for $Pe = 125$, but we observe similar features for 50 $< Pe <$ 200 (data not shown).
In addition to the features already discussed (i.e., preceding density wave and trailing wake), the experimental data shows the emergence of clear banding structure trailing the particle. The average distance between the peaks of neighboring bands is 0.88-0.92 $\sigma$. The absence of comparably prominent and sharp banding preceding the particle suggests that the motion of the probe induces a partial rearrangement of the hexatic fluid. Furthermore, the inter band spacing (measured at 3$\sigma$ behind the probe particle) is intermediate between the spacing expected for a square lattice ($1\sigma$) and that for a triangular lattice ($\sqrt{3}\sigma/2\approx 0.866\sigma$).

To parse the roles of contributing interactions, 
we performed simulations for three different constant probe velocities: $\sigma V/D_{ii} =$ 20, 30, and 50.  This ratio differs from the P\'eclet number in that $D_{ii}$ is not the diffusion coefficient at infinite dilution.  Empirically, we found that the simulations with $\sigma V/D_{ii} = 20$ gave the best correspondence to the experiments at $Pe = 125$, and we show results from those simulations unless otherwise indicated.  We focus on this drag speed because the slower dragging minimizes the artifacts associated with periodic boundary conditions by allowing the system to equilibrate within a single periodic unit of the simulation.  The need to rescale the probe speed likely arises from a number of factors: the simulation is truly 2D, the simulation box is relatively small in size, and the form of the diffusion tensor is extrapolated from lower densities.

Fig.\ \ref{g2SimulExptSmall} shows the pair correlation function for distance ranges well within the periodic boundary conditions.  We see that the simulations reproduce the leading density wave and trailing wake, which suggests that they capture the essential physics of the system. To assess the importance of the hydrodynamic interactions (HI), we compare simulation results with and without HI (Fig.\ \ref{g2SimulExptSmall}c). Without HI we see that the wake still exists, consistent with the observation of similar patterns in lattice Monte Carlo simulations that do not explicitly consider frictional forces \cite{mejia2010bias}.  Quantitatively, the wake is dramatically reduced in extent without HI.  However, if we increase the speed of the driven particle to restore $\sigma V/D_{ii}$ (Fig.\ \ref{g2SimulExptSmall}d), the pair distribution again resembles closely that with full HI.  This comparison between simulations performed in a consistent fashion clarifies the role of the fluid despite potential artifacts arising from the boundary conditions.

\subsection{Individual displacement chains and hexatic order}

\begin{figure}
\includegraphics[width=0.5\textwidth]{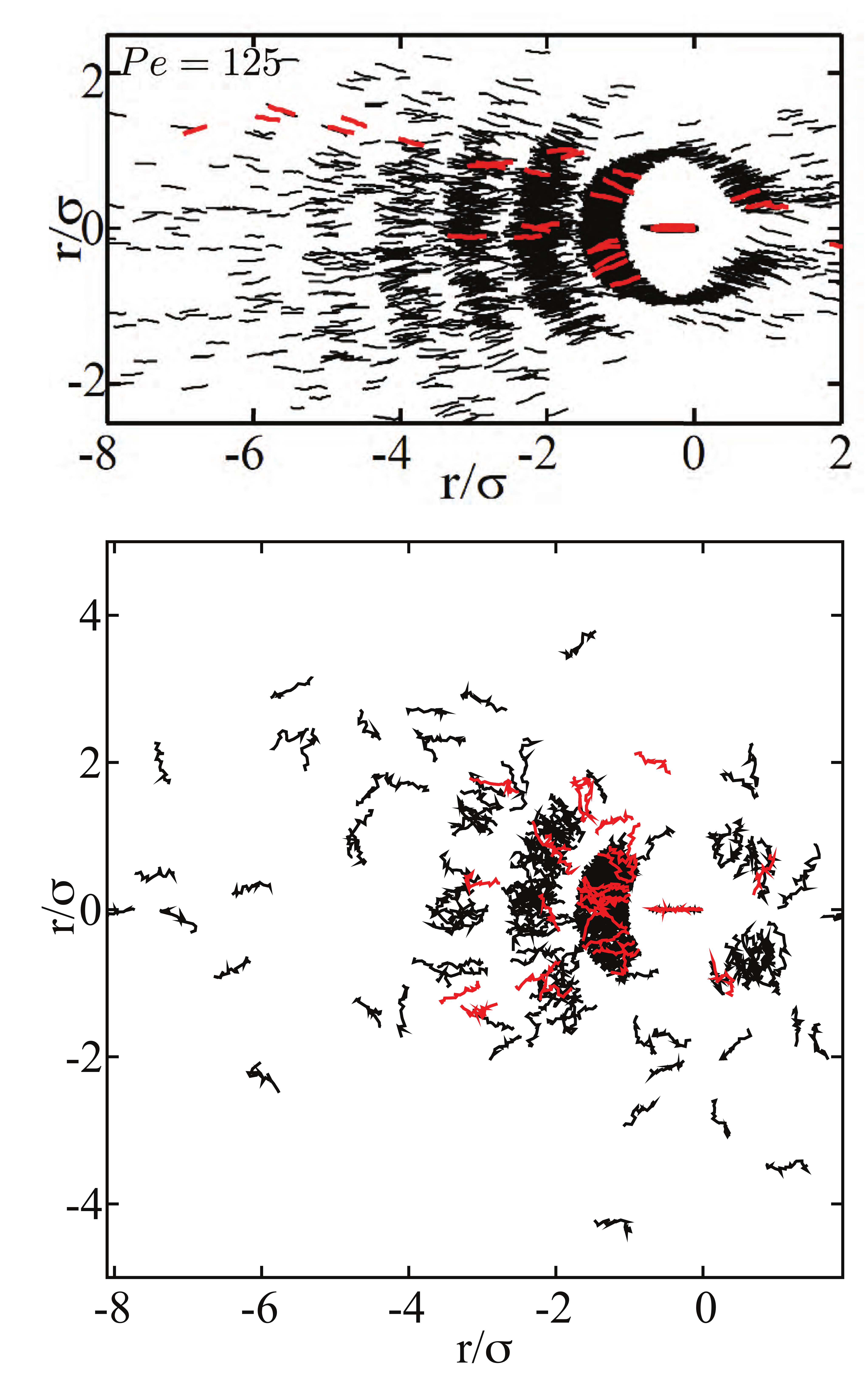}
\caption{
\label{dispChain}
(Color online)
Displacement chains of the bath induced by a probe particle moving at a constant relative velocity ($V$) in experiment (top) and SD simulations (bottom).  A collection of particles forms a displacement chain if each particle's successive displacements are greater than 0.5$d$, where $d$ is the displacement of the probe particle in the dragged frame of reference. The figure depicts a superposition of all displacement chains observed (black) and a sample of representative displacement chains (gray in print or red online).
}
\end{figure}

\begin{figure}
\includegraphics[width=0.5\textwidth]{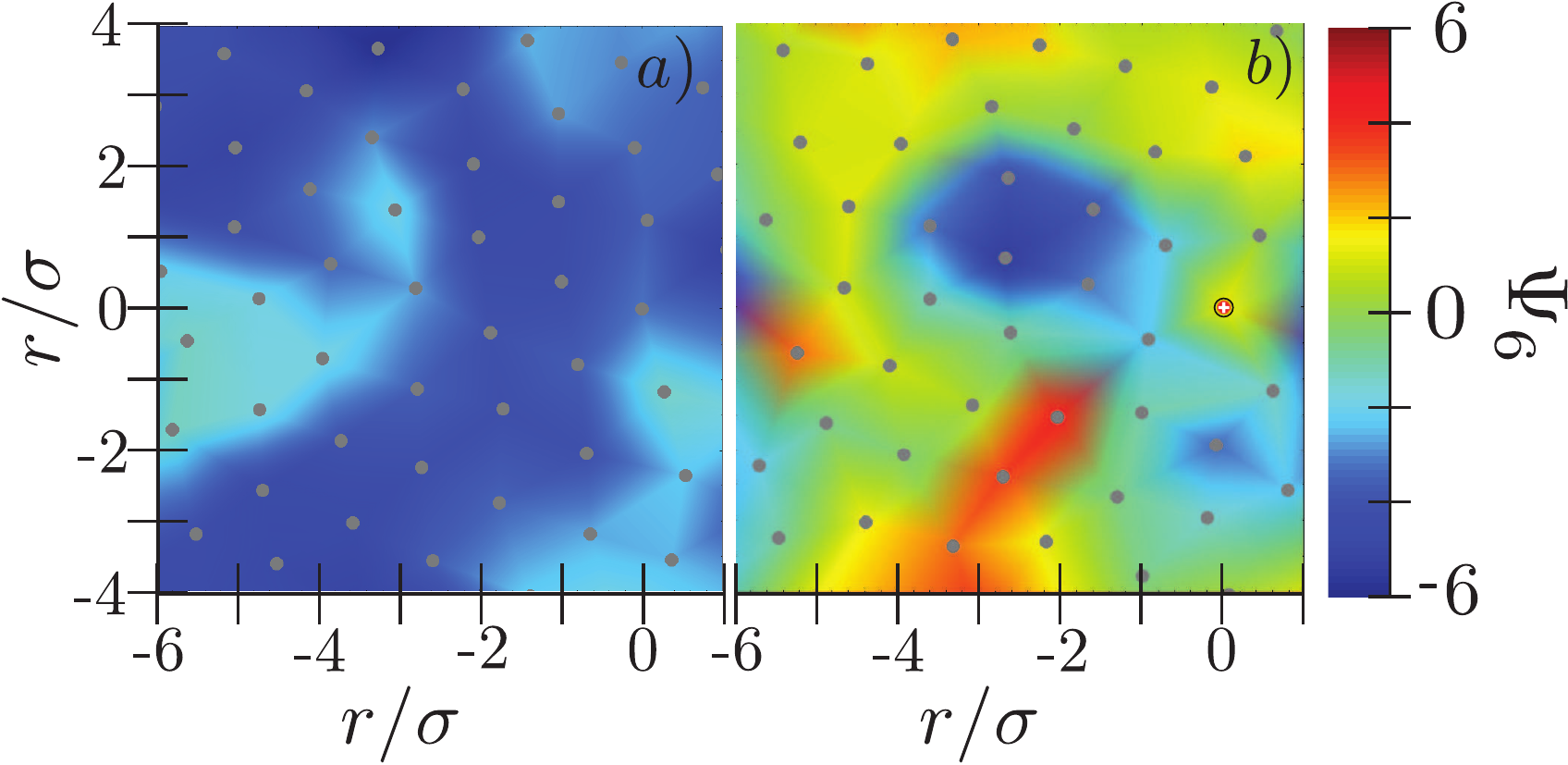}
\caption{
\label{Psi6-Sim}
(Color online)
Bond orientational order parameter, $\Psi_6$, for simulations under equilibrium (left) and constant dragging steady state conditions (right).  The drag direction is to the left and the probe is indicated in red with a white cross at (0,0). }
\end{figure}

As discussed in the Introduction, our goal is to understand how the collective response of the colloid fluid arises from the dynamics of individual particle motions. In Fig.\ \ref{dispChain} we plot the displacements of individual colloids that move more than a threshold amount. Superimposing many such displacements from different times, with the probe particle moving over the same range, reveals a structure consistent with the density wave and wake (black) up to 2 particle diameters from the probe. A representative chain of displacements from a small time interval of the experiment is highlighted in red.  We see that it is quite long, spanning 7 particles. Clearly, the perturbation propagates via colloid-colloid interactions in a highly anisotropic fashion.  Displacements chains are also observed in our simulations extending up to $4\sigma$. Similar behavior has been previously reported in granular simulations \cite{Reichardt2010}
\begin{figure}[t!]
\includegraphics[width=0.5\textwidth]{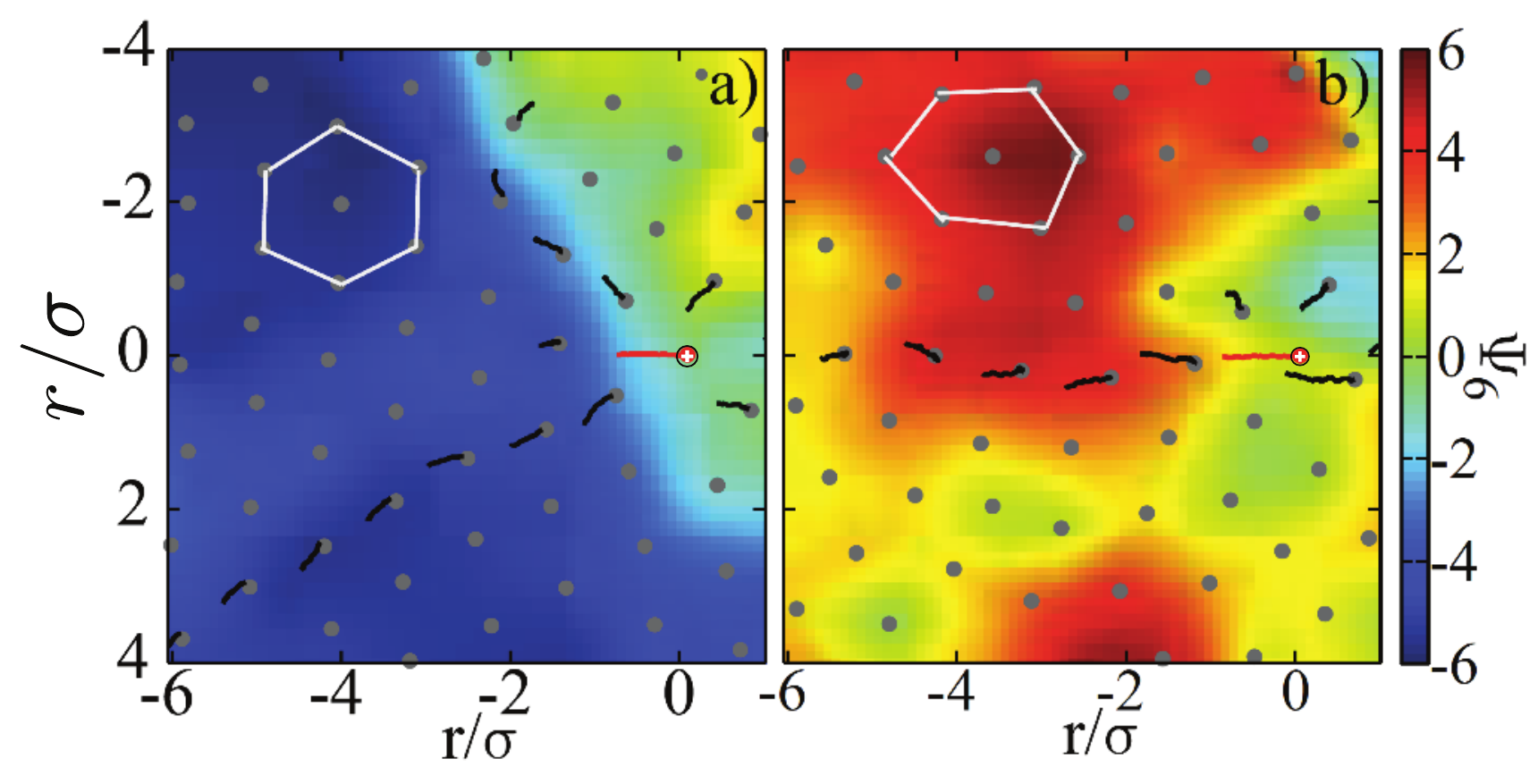}
\caption{(Color online)
Relation between displacement chains and hexatic order.  Representative bath particle displacements over a time interval of approximately 10s are shown in black with the probe particle in red with a white cross at (0,0). The color scale indicates the bond orientational order parameter, $\Psi_6$, calculated for each particle indicated in gray.  White hexagons illustrate the fact that the sign of $\Psi_6$ reports on the orientation of the hexatic structure relative to the direction of motion of the probe particle.\label{Psi6}
}
\end{figure}

At the densities of our experiments the colloid suspension is in the hexatic phase. The extent and orientation of the structural order around a particle can be quantified by the 6-fold bond orientation parameter $\Psi_6=\textrm{Re}\{\sum_{j=1}^6 \textrm{exp}(i\cdot 6\theta_j)\}$.  Here, $\theta_j$ is the angle between a fixed axis and the line connecting the centers of the particle of interest and its $j$-th nearest neighbor. Within regular hexagonal domains, the value of $\Psi_6$ is constant. Changes in the value of $\Psi_6$ correspond to grain boundaries and transitions to disordered domains. Fig.\ \ref{Psi6-Sim} highlights typical bond orientational configurations obtained from simulation data under equilibrium and constant dragging conditions. As expected, in equilibrium we observe bond orientational order on length scales comparable to the simulation cell size. For the constant dragging perturbation, the forced motion of the probe particle distorts the hexatic order through repeated cage breaking ultimately leading to the formation of extended grain boundaries and disordered regions. 

We would like to determine if the size and orientation of the displacement chains is determined by the local structure.  Fig.\ \ref{Psi6} highlights typical situations. The sign of $\Psi_6$ characterizes the orientation of the hexatic structure relative to the direction of motion of the probe particle (compare the orientation of the white hexagons with the color scale in Fig.\ \ref{Psi6}). In each case, the displacement chain propagates according to the orientation of the local structure; displacement chains form inline with the probe particle when $\Psi_6 > 0$ and bifurcate when $\Psi_6< 0$.  However, since the size of ordered domains is typically less than $10\sigma$ across, the displacement chains can span multiple domains and hence are sensitive to grain boundaries. The upper displacement chain in Fig.\ \ref{Psi6}a follows a grain boundary, indicating that both local order and defects influence the bath response. 
The approximate angular spread ($\sim 60^\circ$) of the superposition in Fig.\ \ref{dispChain} is consistent with the largest angular spread occurring when the probe bisects the hexatic ordering, i.e., when $\Psi_6 = -6$ as shown in Fig.6a.
Orientational effects in driving particles through larger ordered domains have been investigated previously \cite{Reichardt2004}.

\subsection{Dipolar-like flow}

In the Stokeslet approximation, the perturbation caused by the probe particle induces a dipole-like flow in the supporting liquid. 
This prediction is valid for the far-field flow in the limit of a dilute suspension of point particles. In contrast to these assumptions, in our system the particle density is high, the experimental ratio of particle diameter to confinement thickness is 0.8,  and the far-field flow limit is not experimentally accessible owing to spatial correlations present in the hexatic phase that are comparable to the longest measurable length scales.
Furthermore, given the qualitative connections of the dense colloid packing to granular systems \cite{liu1995force, candelier2009creep, harich2011intruder}, we expect the observed displacement chains (Fig.\ \ref{dispChain}) to introduce an anisotropic component to the flow field.
Nevertheless, one might expect that qualitative features of the dipolar flow of the liquid to persist. Specifically, we expect the closed loop circulation pattern to persist, but not the algebraic form of the fall off of velocity with the distance from the source. We attribute the change in spatial fall off of the velocity field to interferences generated by scattering from the densely packed colloid particles and the walls, drawing on our experience with calculations using the method of reflections \cite{xinliang2005Reflections}. This method represents the influence of the boundaries on the velocity field generated by the moving particle as the superposition of the velocity components reflected from each boundary encountered by the initially generated velocity field. The reflections are considered to be coherent, so the superposition of the velocity components has both positive and negative interferences. In a system with many reflecting boundaries, as in the case of densely packed colloids studied here, these interferences can greatly change the symmetry and spatial falloff of the velocity field. 

\begin{figure}[h!]
\includegraphics[width=0.5\textwidth]{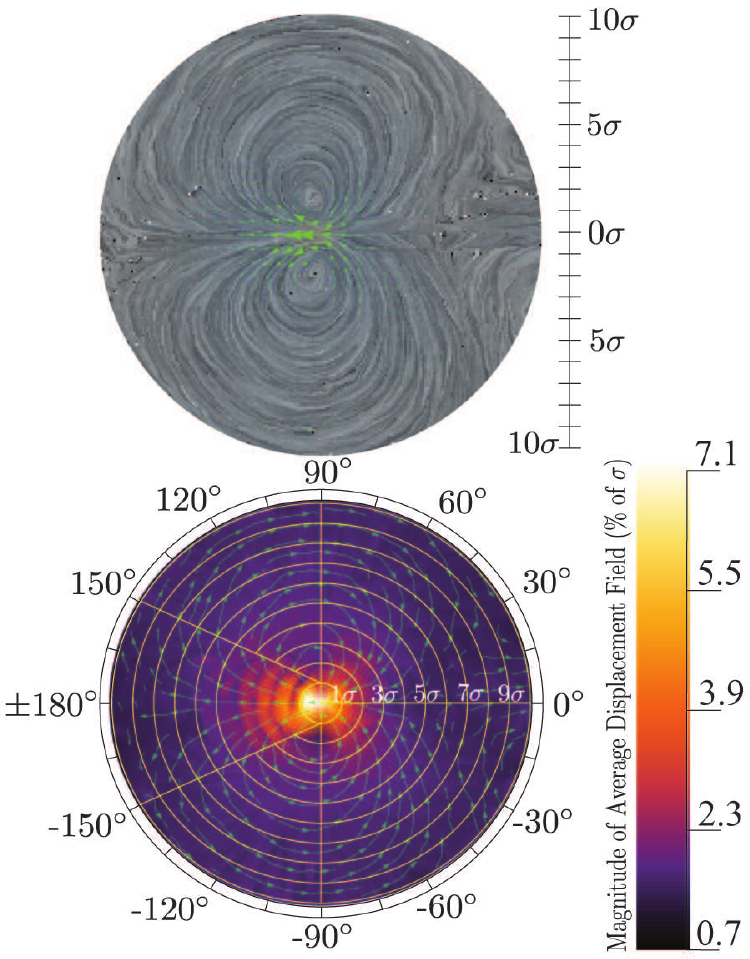}
\caption{
\label{dipolar}
(Color online)
Dipolar-like flow fields around a probe particle moving at constant velocity for $Pe=190$. The dragging direction is to the left and the field of view has a diameter of $20\sigma$. The interpolated vector field is obtained from the frame to frame displacements of every particle in the field of view using a combination of spatial binning and time averaging   (see text). Fig.\ \ref{dipolar}a shows a sample of the interpolated vectors (green arrows) of the averaged displacement field and the associated line integral convolution (gray) indicating the extent of weak but coherent dipolar flow. Fig.\ \ref{dipolar}b shows streamlines (green arrows) obtained from the interpolated vector field and the magnitude of the average displacement field (gradient). 
}
\end{figure}

\begin{figure}[!hbt]
\includegraphics[width=0.5\textwidth]{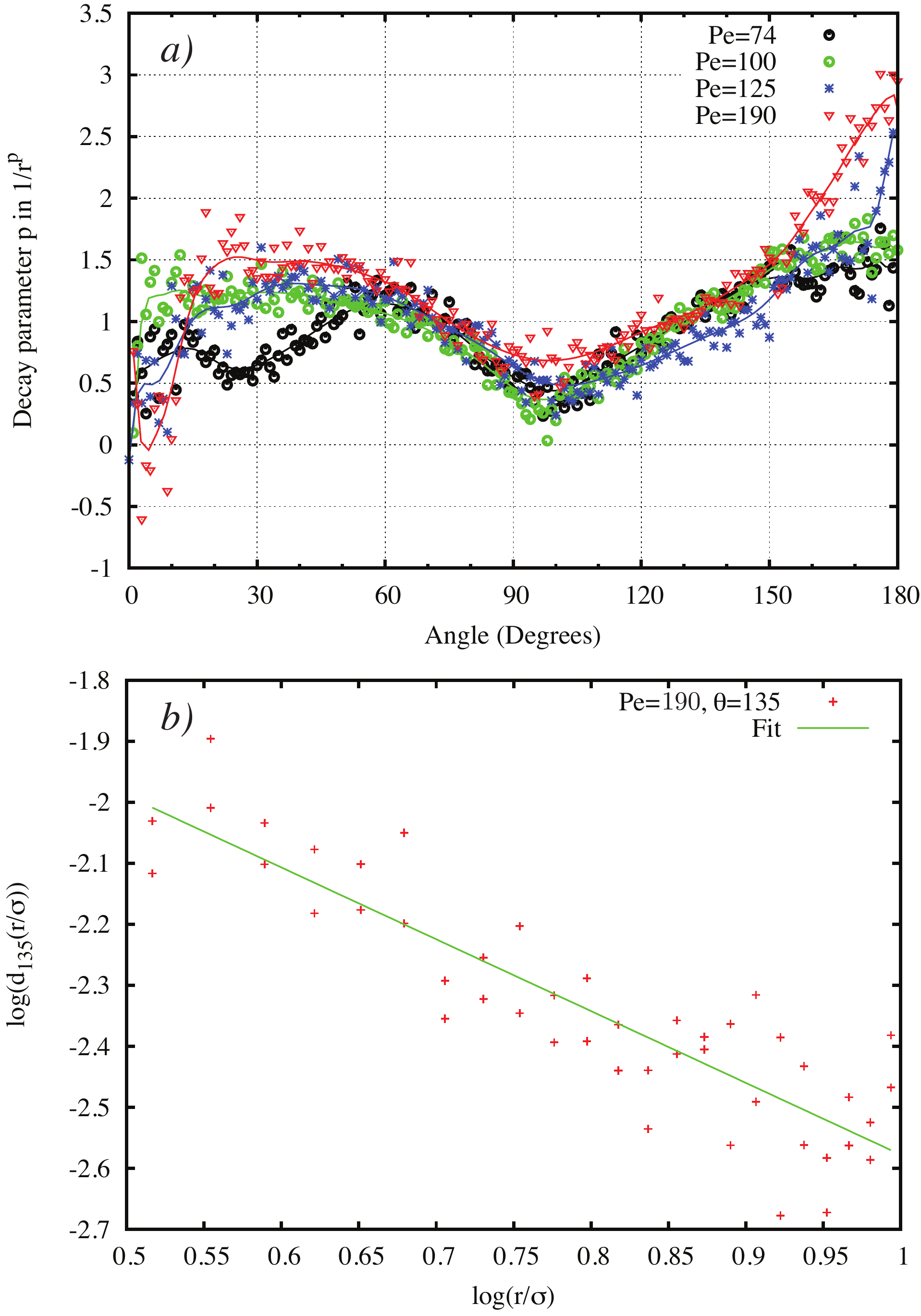} 
\caption{
\label{thetaDepDecay} 
(Color online)
Decay parameter of the dipolar-like flow field. (a) Angle dependence.  The angle corresponding to motion of the probe particle is 180 degrees. 
(b) Typical linear fit of log$_{10}$-log$_{10}$ data for determining the angle dependent decay parameter. The log$_{10}$-log$_{10}$ data were fitted to $f(x)= b-mx$ and the fitted parameters were $m=1.17 \pm 0.09$ and $b=-1.40\pm0.07$. The fitted negative slope, $m$, corresponds to a single decay parameter quoted in (a). The angle shown corresponds to 135 degrees. The uncertainties quoted are asymptotic errors in a least squares fit. } 
\end{figure}

We recover the dipolar-like flow shown in Fig.\ \ref{dipolar} through a combination of spatial binning and time averaging. To construct the field from experimentally measured particle trajectories, the frame-to-frame displacement ($\Delta_x, \Delta_y)$ for each particle was computed in the dragged frame of reference:
\begin{equation}(\Delta_x, \Delta_y) \equiv \Big(x(t+\delta t) - x(t), y(t+\delta t) - y(t)\Big),\end{equation}
where $\delta t$ is the time between frames, and $x(t)$ is the position of particle $i$ at time $t$.
The precise components of the drag speed depend on the angular position of the camera relative to the stage. This introduces an additional source of uncertainty, estimated at $<3\%$ in each component, in determining the displacement field components in the dragged frame of reference. We choose the drag velocity within the uncertainty range that maximizes the spatial extent of the dipolar-like flow. This procedure is justified by the consistency of the optimization process in obtaining far field dipolar-like flow, and the reproducibility of the structural form of the angle dependent decay in Fig.\ \ref{thetaDepDecay}.

The displacements were spatially binned based on the angle and radial distance relative to the probe particle. This was done for every frame of the video, after which each bin was time averaged. Only spatial bins that had more than 30 samples were used to generate an interpolated vector field. The threshold of 30 samples was sufficiently high to ensure that incoherent thermal motions were averaged out while ensuring that the whole vector field, including regions of vanishing flow, was adequately sampled. Fig.\ \ref{dipolar}a shows the interpolated vector field, depicted by green arrows, superimposed on a line integral convolution (LIC) flow visualization \cite{cabral1993imaging} which emphasizes the weak but coherent dipolar-like flow observed out to ten diameters away from the probe particle. Fig.\ \ref{dipolar}b shows streamlines (green arrows) and the magnitude of the average displacement field (gradient). To the best of our knowledge, this is the first experimental observation of dipolar-like flow in q2D colloid suspensions, and, more generally, the only quantitative analysis of such a feature in a complex fluid.

As already noted, the dipolar-like flow that is generated in the supporting liquid by the driven particle is expected from a simplified analysis of the hydrodynamic behavior of a q2D colloid system wherein the driven particle is treated as a force monopole in an incompressible liquid confined between two plates \cite{dufresne2001brownian,liron1976stokes}. As each colloid particle moves in the plane parallel to the walls it changes the momentum and mass density of both the colloid particle and the supporting liquid \cite{diamant2009hydrodynamic}. The momentum of the supporting liquid is transferred to the confining walls, and is not conserved. The mass of the colloid particle and that of the supporting liquid are conserved, and the mass perturbation associated with particle motion dominates the flow pattern at distances larger than the thickness of the sample. To leading order in the ratio of particle radius to distance from the particle, the mass perturbation can be represented as a mass dipole, which in q2D creates a flow velocity proportional to $r^{-2}$ at large distances from the driven particle. It is this functional form that defines the long-range flow pattern in q2D confinement. On a time scale short relative to the time between colloid-colloid collisions, the colloid particles act as reporters of the local fluid velocity, rendering the dipolar-like flow observable on a length scale comparable to multiple colloid-colloid separations. 
As already noted, in a dense colloid suspension we expect deviation from the $r^{-2}$ behavior because of interference with the velocity field generated by the zero slip boundary condition on each colloid particle with nonzero volume.  

\begin{figure}[!hbt]
\includegraphics[width=0.5\textwidth]{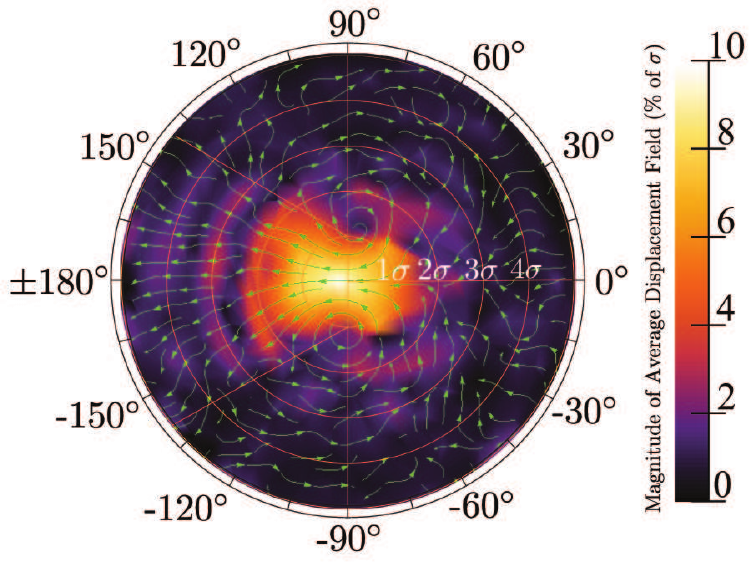}
\caption{
\label{dipolarsim} 
(Color online)
Dipolar-like flow fields observed in simulations. The field of view has a diameter $10\sigma$ which spans a single simulation cell. The color scale corresponds to the magnitude of the displacement field and the streamlines are depicted by green arrows. }
\end{figure}

Given the anisotropic structure of both the average colloid packing (Fig.\ \ref{g2Pe20Large}) and the intermittent colloid structure (Fig.\ \ref{dispChain}), one should expect that colloid displacements reflect not only the average displacement of the underlying fluid, but also the observed structural anisotropy. 
Keeping in mind the roles of (i) colloid-colloid interactions, (ii) hydrodynamic screening due to the high density, and (iii) nonzero size of the probe particle, we determine the decay parameter $p$ in $1/r^p$ as the slope of a linear fit of $\log_{10}[d_\theta(r)]$ as a function of $\log_{10}(r)$ where $d_\theta(r)$ is the magnitude of the  displacement vector of the dipolar-like flow field at a radial distance $r$ from the probe particle and fixed angle $\theta$. The fitting region was $r\in(3\sigma,10\sigma)$.  As shown in Fig.\ \ref{thetaDepDecay}a, the decay parameter $p$ exhibits a strong dependence on $\theta$. The observed angle dependence may be qualitatively understood as a manifestation of the nontrivial angle dependent structure of the hexatic fluid, as characterized by the pair correlation function, and the corresponding colloid-colloid interactions. We note that some quantitive aspects of Fig.\ \ref{dipolar} and Fig.\ \ref{thetaDepDecay}a (e.g., spatial extent of coherent dipolar-like flow beyond $5\sigma$ in Fig.\ \ref{dipolar}, and angular location of minimum and quantitative values of the weak decay parameters in Fig.\ \ref{thetaDepDecay}a) are sensitive to the components of the drag speed within our uncertainty range but the qualitative features discussed above are not.

The modulation of the displacement field preceding the probe particle,  shown in Fig.\ \ref{dipolar}, exhibits peaks corresponding to integer particle diameters and a cone consistent with the angular distribution of displacement chains shown in Fig.\ \ref{dispChain}. The anisotropic displacement chains, which are a direct result of correlated colloid-colloid collisions, perturb the long range dipolar flow field. Qualitatively similar flow fields were observed in driven granular systems \cite{candelier2009creep, harich2011intruder}. Our simulations, regardless of whether HI are included, show dipolar-like flow fields (Fig.\ \ref{dipolarsim}) that are similar to the experimental results (Fig.\
\ref{dipolar}). We obtain the displacement fields shown in Fig.\ \ref{dipolarsim} analogously to the experimental system. In the experiment, the displacements $(\Delta_x, \Delta_y)$ were computed using adjacent frames; in the simulation, the displacements were computed over the number of simulation time steps required for the probe particle to move a distance of $\sigma/10$. In addition to the dipolar-like flow fields, the displacement field also shows velocity modulation, preceding the probe, occurring at integer particle diameters as in the experiment. However, the decay is clearly faster than in Figs.\ \ref{dipolar} and \ref{thetaDepDecay}.  Since we obtain similar results even for very large systems without HI (data not shown), we do not believe this is a boundary artifact.  Rather, the simulation results suggest that the assumed form for the HI, while adequate for elucidating qualitative features of the observed dipolar-like flow fields, becomes inappropriate for quantitative analysis of the decay parameters.
At high densities, an alternative scaling and parameterization may be required to capture the  interplay between granular and liquid components in the colloid fluid, which is dominated by near-field effects.

While a first-principles analysis of experiment and simulation is not possible, we can relate the qualitative features observed in Fig.\ \ref{thetaDepDecay} to the phenomenology observed in the dipolar-like displacement fields in both experiment (Fig.\ \ref{dipolar}) and simulation (Fig.\ \ref{dipolarsim}). As in Fig.\ \ref{thetaDepDecay}, we adopt the convention that the driving direction is $180^\circ$ and the system is approximately symmetric with respect to the drag direction. Consider the following regions: 

\begin{itemize}
\item Region I, $|\theta| \in (0^\circ,20^\circ)$: This region corresponds to the wake. Slow decay is observed suggesting that the dipolar flow is dominated by the slow structural relaxation of the wake. Although the fitting region ($>3\sigma$) lies outside the wake (Fig.\ \ref{g2Pe20Large} and Fig.\ \ref{g2SimulExptSmall}) the sampling is poorer in this region than in others.
\item Region II, $|\theta| \in (20^\circ,60^\circ)$: This region corresponds to dipolar-like circulation currents that flow into the wake. Decay parameters with typical values $p=1.5$ are observed which are between those expected for a point force in 3D ($p=1$) and the point stokeslet limit in q2D ($p=2$). 
\item Region III, $|\theta| \in (60^\circ,150^\circ)$: This region spans the location of the vortex. In each of the drag speeds consider in the experiment, the minima in angle dependent decay at $\sim 100^\circ$ corresponds to the angular position of the vortex. Comparing the point stokeslet approximation and the experimental conditions, we hypothesize that the non-zero size of the probe particle plays a significant role in forming the vortex. Additionally, we note that both simulation and experiment show characteristic angle dependent anisotropy in the displacement field, with the location of the vortex corresponding to a steep decline in the displacement magnitude.   
\item Region IV, $|\theta| \in (150^\circ,180^\circ)$: This region spans the velocity modulation of the dipolar-like field directly caused by the colloid-colloid collisions. The faster decay parameters ($p > 2$) show that momentum transfer through colloid-colloid collisions decays more quickly than hydrodynamic interactions, further suggesting that the long range, coherent dipolar-like flow fields arise from HI.
\end{itemize}

\section{Conclusions}

Our study establishes the phenomenology of response to a localized perturbation in a dense q2D colloid suspension and allows it to be related to individual particle-particle collisions.
The trapped particle motion generates a density wave that precedes it and a void in its wake.  These ensemble-averaged features have been widely observed in experiments in unconstrained (3D) colloid systems \cite{meyer2006laser,sriram2010active} and in 2D simulations \cite{squires2008nonlinear}.  We have demonstrated that these features are preserved in a q2D geometry as well and have shown how they arise from individual displacement chains. 

Our study provides insight into the range of validity of the Smoluchowski equation treatment of the response of a colloid suspension to the driven motion of one particle \cite{khair2006single}. This approach neglects multiparticle (more than two) hydrodynamic interactions and is thereby limited to low density when used to describe a three-dimensional suspension. It should remain valid to higher densities in our case because the three particle hydrodynamic interaction vanishes in a q2D colloid suspension at the Stokeslet level of approximation, 
and our simulations did indeed capture the leading density wave and trailing wake well.

That said, the overall deviation of the decay of the dipolar-like field from the point Stokeslet approximation makes clear the importance of the fact that the colloids are comparable in size to the separation between the boundaries.  The striking angular dependence of the decay reflects the structure of the surrounding colloids, which evolves through individual displacement chains arising from the perturbation.
Incorporating the anisotropic propagation of forces and these memory effects will be important for theories and simulations that seek to describe the response of such complex fluids fully.  Although doing so will be challenging, success  would have repercussions not only for colloid suspensions but for a much broader range of fluids, including granular ones.

We will report complementary results from studies of relaxation dynamics associated with nonequilibrium excitations that create density waves, wakes, and persistent void areas in a separate publication. 

\section{Acknowledgements}

We thank Binhua Lin for assistance in the design and fabrication of the sample chamber.  JZT acknowledges a summer research fellowship from the James Franck Institute. We acknowledge financial and central facilities assistance of the University of Chicago Materials Research Science and Engineering Center, supported by the National Science Foundation (NSF DMR-MRSEC 0820054).


\end{document}